# OD + CO → D + CO$_2$ Branching Kinetics Probed with Time-Resolved Frequency Comb Spectroscopy

**Authors:** Thinh Q. Bui*, Bryce J. Bjork, P. Bryan Changala, Oliver H. Heckl, Ben Spaun, Jun Ye*

**Affiliation:**

JILA, National Institute of Standards and Technology and University of Colorado, Department of Physics, University of Colorado, Boulder, CO 80309, USA

*Correspondence to: thbu8553@jila.colorado.edu, ye@jila.colorado.edu

**Abstract**

Time-resolved direct frequency comb spectroscopy (TRFCS) was used to study the kinetics of the deuterated analogue of the OH+CO→H+CO$_2$ reaction, which is important for atmospheric and combustion chemistry. Complementing our recent work on quantifying the formation rate of the transient *trans*-DOCO radical, we report measurements of the kinetics of the activated product channel, D+CO$_2$, at room temperature. Simultaneous measurements of the time-dependence of OD and CO$_2$ concentrations allowed us to directly determine the activated products' formation rate, branching yield, and dependences on pressure and bath gas. Together with the *trans*-DOCO formation rate, these new measurements provide absolute yields of branching channels for both products of OD+CO in the low-pressure limit.

1. **Introduction**

The reaction,

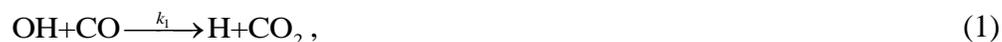

$$\mathrm{OH+CO} \xrightarrow{k_1} \mathrm{H+CO_2},  \qquad (1)$$

has served as a benchmark system for kinetics and dynamics studies of complex-forming, bimolecular reactions for the past four decades because of its importance in atmospheric and combustion chemistry.[1] On Earth, CO is a byproduct of fossil fuel burning and hydrocarbon oxidation and acts as a global sink for OH radicals in the free troposphere. In fossil fuel combustion, reaction (1) is the main oxidation step to convert CO to $CO_2$. Based on a recent proposal by Boxe et al.[2], the OH+CO reaction may also play a significant role in explaining the $CO_2$ budget on Mars: reactions involving the long-lived HOCO radical intermediate may be a key catalytic source of $CO_2$ production.

The OH+CO reaction is given by the following elementary reaction steps:

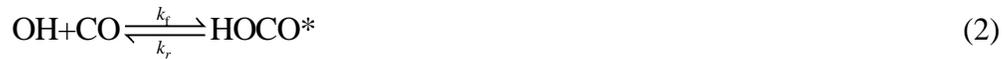

$$OH + CO \underset{k_r}{\overset{k_f}{\rightleftharpoons}} HOCO^* \quad (2)$$

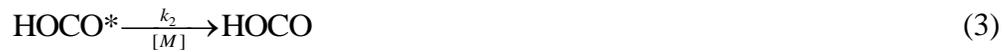

$$HOCO^* \xrightarrow{\frac{k_2}{[M]}} HOCO \quad (3)$$

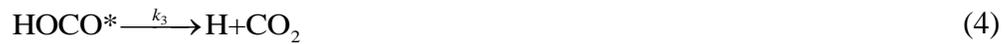

$$HOCO^* \xrightarrow{k_3} H + CO_2 \quad (4)$$

The OH+CO proceeds to first form the (vibrationally) energized HOCO*, which can (i) dissociate back to OH+CO, (ii) relax to ground state HOCO by third-body collisions with bath gas M, at a rate coefficient $k_{1a}$, and/or (iii) decompose to produce the activated products, H+$CO_2$, at a rate coefficient $k_{1b}$. The formation of the HOCO radical complex leads to the observed non-Arrhenius temperature and strong pressure dependences of the rate coefficient.[3-7] Based on this scheme, (ii) HOCO and (iii) H+$CO_2$ formation dominate in the high- and low-pressure limits, respectively. The overall reaction rate, $k_1$, is simply described by an effective bimolecular rate constant $k_1([M],T) = k_{1a}([M],T) + k_{1b}([M],T)$.

The temperature and pressure dependences of the OH+CO rate coefficients have been studied extensively.[3-15] Purely *ab initio* methods involving master equations can provide estimates for thermal rate coefficients of complex-forming, pressure-dependent reactions; however, such endeavors are often hindered by incomplete accounting of collision and energy transfer dynamics for activation and stabilization of intermediate complexes.[16, 17] Moreover, the rate coefficients are particularly sensitive to the collisional and energy transfer parameters in the low-pressure limit and fall-off regions (intermediate pressure range between the low- and high-pressure limits). In the case of the OH+CO reaction, these problems have persisted for over four decades. The underlying dynamics involving the HOCO radical intermediate have previously been understood only in terms of empirical fit models[4, 6, 9, 10, 18] and master equation calculations[1, 19, 20] used to fit the measured decay rate coefficients of OH in the presence of CO, $k_1$.[17] Experimentally, quantitative kinetic measurements of pressure-dependent branching, i.e. stabilization to HOCO (eq. 3) and activation to H+$CO_2$ (eq. 4), are necessary prerequisites. Yet, the HOCO intermediate has eluded detection in thermal environments until recent experimental demonstration by Bjork *et al*.[21] on the first direct measurements of the formation rate ($k_{1a}$) of the deuterated analogue, *trans*-DOCO, in the OD+CO reaction. This study complements the Bjork *et al*. work in providing the direct measurements of the formation rate ($k_{1b}$) of the activated products, D+$CO_2$, at thermal conditions using frequency comb spectroscopy. Together, the goal of these studies is to provide quantitatively mechanistic details of the OD+CO reaction in the low-pressure limit, a good test case for studying effects of collisional energy transfer on rate coefficients for this important complex-forming reaction.

2. **Experimental**

Time-resolved frequency comb spectroscopy (TRFCS) has been developed for applications of spectroscopy and dynamics of transient radicals.[22, 23] Relevant details for this experiment are found in our previous publication.[21] A high repetition rate ($f_{rep}$ = 136 MHz) mode-locked femtosecond fiber laser synchronously pumps an optical parametric oscillator (OPO) to produce the mid-IR comb light spanning from 3 to 5 μm.[24] The mid-IR comb light is injected into a high-finesse cavity, whose free spectral range (FSR) is matched and locked to 2×$f_{rep}$. The transmitted light is spatially dispersed by a virtually-imaged phased array (VIPA) etalon and a grating combination, which is then imaged onto an InSb camera.[25] Absorption spectra as a function of time (limited by camera integration time ≥ 10 μs) are constructed from the camera images and fitted to known line intensities of reference molecular spectra to obtain absolute concentrations. In general, this technique provides a unique combination of broad bandwidth spectroscopy, high sensitivity, high spectral resolution and microseconds time resolution for simultaneous detection of a number of key species in the reaction. The main modification to the previous TRFCS instrument is the use of high finesse mirrors centered at 3.92 μm from Lohnstar Optics.[26] Mid-IR mirror finesse measurements were conducted in the same manner described in Cole et al.[27] These mirrors provide both a high finesse (F ≈ 5600) and a large bandwidth (> 400 nm), as shown in Fig. 1. The large bandwidth provides access to many molecular species relevant to the OD+CO reaction including *trans*-DOCO, $CO_2$, $DO_2$, $D_2O$, and OD(v=0-4). All of these have been measured in this experiment except for OD(v=3, 4).

The OD+CO reaction was studied in a flow cell under reaction conditions kept nearly the same as those described in Bjork *et al*. Major sources of systematic error have been characterized under these conditions, including rate constant dependences on camera integration time, vibrationally excited OD contributions, and ozone and $D_2$ gas concentrations. Here, we will provide only a brief summary of the experimental procedures. O($^1$D) atoms are first generated from photolysis of O$_3$ at 266nm (35 mJ/pulse) from a frequency-quadrupled Nd:YAG laser. Each photolysis pulse dissociates about 15% of the ozone to form O($^1$D) and O$_2$. In the presence of D$_2$, O($^1$D) + D$_2$ produces energized OD(v=0 - 4) with an inverted population peaking at v=2 and v=3.[28] High CO concentrations (> 3.5×10$^{17}$ molecules cm$^{-3}$) were maintained for the purpose of ensuring the low densities and short lifetimes of OD(v>0). The OD(v=1) + CO quenching rate constant has been previously measured to be (3.3±0.2)×10$^{-13}$ cm$^3$ molecules$^{-1}$ s$^{-1}$.[21] Using this value, the maximum lifetime (1/e) of OD(v=1) at our conditions is 8.7 μs. Maintaining a concentration of high CO ensures that the contribution from vibrationally excited OD on the uncertainties of $k_{1b}$ is less than 10% based on both the lifetime and abundance detection sensitivity of OD(v>0).

Molecular densities of D$_2$, CO and N$_2$ were controlled and monitored with mass flow regulators and meters. For all experiments, the O$_3$ concentration was fixed at 2×10$^{15}$ molecules cm$^{-3}$ and monitored *in situ* by UV absorption spectroscopy at 270 nm. The D$_2$ concentration was also kept constant at 7.4×10$^{16}$ molecules cm$^{-3}$. By controlling the partial pressures of He, CO, and N$_2$ gases, the experimental total pressures were varied from 40 to 120 torr.

3. **Results and discussion**

Absorption spectra covering ≈60 cm$^{-1}$ of bandwidth centered at 2420 cm$^{-1}$ (≈4.13 μm) were recorded with a varying time delay from the t=0 photolysis pulse. Each spectrum was normalized to a reference spectrum acquired before the photolysis and fitted to determine the time-dependent concentrations. For all experiments, the camera integration time was fixed at 100 μs. Fig. 2A show representative snapshots of measured and simulated spectra at time delays 25 and 1000 μs after the photolysis pulse. The R(76) transition of CO$_2$ near 2390.522 cm$^{-1}$ (line intensity S=4.140×10$^-$

$^{22}$ cm molecule$^{-1}$) and OD transition near 2435.5 cm$^{-1}$ (S=1.64×10$^{-21}$ cm molecule$^{-1}$) are the strongest absorption features. The time-dependent curves in Fig. 2B were obtained from fitting integrated areas for both molecules. The OD line intensities are determined from transition dipole moments calculated using the empirical potential energy and dipole moment surfaces reported by Nesbitt and coworkers.[29, 30] The CO$_2$ line intensities were taken from the HITRAN 2012 database.[31]

The rate coefficients for the D+CO$_2$ channel, $k_{1b}$([M],T), were determined from simultaneous measurements of time-dependent [CO$_2$]($t$) and [OD]($t$). $k_{1b}$ may be bimolecular (independent of pressure) or termolecular, depending on whether the conditions are at the low, intermediate, or high-pressure limits.[4, 6, 18] We evaluated these scenarios by measuring the dependences of the effective bimolecular rate constant on the concentrations CO, N$_2$, and He.

The time-dependent CO$_2$ formation rate is given by the rate law

$$\frac{d[CO_2]}{dt} = k_{1b}[CO][OD](t), \qquad (5)$$

where [OD]($t$) refers to time-dependent concentration of OD in the vibrational ground state. Contrary to DOCO, CO$_2$ does not have a large loss channel, e.g. reactions with O$_3$, on the time scale (<1 ms) of our rate constant determination. Solving eq. 5 for [CO]($t$) gives

$$[CO_2](t) = k_{1b}[CO]\int_0^t [OD](u)du. \qquad (6)$$

Since [CO] is in large excess and remains constant throughout the reaction, quantifying $k_{1b}$ requires only the time dependences of [CO$_2$] and [OD]. For [OD]($t$), we used derived analytical functions comprised of the sum of boxcar-averaged exponential rise and fall functions. Eq. 6 gives the functional form for fitting [CO$_2$]($t$), which includes the integrated [OD]($t$) over the fitted time window of 0 to 1000 μs. Finally, $k_{1b}$ may have dependences on bath gas and pressure, i.e.,

$$k_{1b} = k_{1b}^{(CO)}[CO] + k_{1b}^{(N2)}[N_2] + k_{1b}^{(He)}[He]. \tag{7}$$

Representative plots of fits to both [CO$_2$]($t$) and [OD]($t$) are shown in Fig. 2B. The boxcar-convolved fits for [CO$_2$]($t$) and [OD]($t$) reveal that multiple time regimes are in effect, which is expected due to the secondary regeneration channels of OD. At our conditions, OD decay is observed to be bi-exponential, with the initial decay (lifetime ≈100 to 300 μs) coming from reactions with CO. The second exponential decay (lifetime ≈1000 μs) occurs approximately after 300 μs and reaches a near steady-state at longer times (t >1000 μs). OD regeneration reactions D+O$_3$→OD+O$_2$ and DOCO+O$_3$→OD+CO+O$_2$ dominate at longer times, consistent with previous observations.[4, 21] Therefore, only the earliest time behavior (<300 μs) captures the initial OD+CO→D+CO$_2$ branching reaction and is used for the analysis of $k_{1b}$.

The bath gas and pressure dependences of the bimolecular rate constant $k_{1b}$ were measured for CO, N$_2$ and He gas. For CO, the range of densities was limited from 3.5×10$^{17}$ to 1.0×10$^{18}$ molecules cm$^{-3}$. Too low of CO densities were avoided due to complications from vibrationally excited OD since CO is an efficient quencher of OD vibration. Too high of CO densities limit the signal-to-noise of OD detection because of O($^1$D)+CO quenching. The effects of vibrationally excited OD in this system have been systematically analyzed previously.[21] For N$_2$ and He gas, the upper limit densities of ≈ 3.5×10$^{18}$ molecules cm$^{-3}$ were dictated by technical limitations: high molecular densities result in too much shock from the photolysis pulse which affects cavity locking stability. The results of the $k_{1b}$ measurements are shown in Fig. 3. Within 1σ measurement uncertainties, $k_{1b}$ was observed to be constant with respect to pressure for all three bath gases. The average $k_{1b}$ displayed values of $5.6(4)\times10^{-14}$, $6.6(8)\times10^{-14}$, and $6.1(4)\times10^{-14}$ for CO, N$_2$, and He, respectively.

The observed $k_{1b}$ results may be rationalized from unimolecular rate theory and the associated Lindemann mechanism common to pressure-dependent reaction kinetics.[4, 6, 18] Starting from processes described in eqs. 2-4 and applying the steady-state approximation (d[DOCO*]/dt = 0), the CO$_2$ formation rate is

$$\frac{d[CO_2]}{dt} = \frac{k_f k_2}{k_r + k_2 + k_3[M]}[OD][CO]. \tag{8}$$

Recasting the first factor in eq. 8 and applying the limits of [M]$\rightarrow$0 and [M]$\rightarrow\infty$ yield the low- and high-pressure limit rate constants given by eqs. 9 and 10, respectively:

$$k_{1b,[M]\rightarrow 0} = \frac{k_f k_2}{k_r + k_2}, \tag{9}$$

$$k_{1b,[M]\rightarrow \infty} = \frac{k_f k_2}{k_3[M]}. \tag{10}$$

Here, $k_{1b}$ is independent of [M] as pressure approaches zero and inversely proportional to [M] at infinite pressure. The experimentally observed pressure-independent behavior of $k_{1b}$ is consistent with predictions by this Lindemann-type mechanism in the low-pressure limit (eqs. 8-9). Similar observations were reported previously from empirical fits to $k_1$ from literature measurements in the low-pressure limit, as discussed in detail by Fulle *et al.*[4] and Golden *et al.*[6] Any apparent curvature in the pressure-dependences would suggest deviations from the Lindemann mechanism or departure from the low-pressure limit and transition into the fall-off region, which we cannot completely rule out upon inspection of Fig. 3.

Based on $k_{1a}$ from Bjork et al. and $k_{1b}$ from this work, we can directly compare our measured values to literature measurements of $k_1$. According to calculations by Weston et al.[19] and the empirically-derived forms of $k_{1a}$ and $k_{1b}$ from Fulle et al., most of the pressure-dependence of $k_1$ comes from $k_{1a}$ in the low-pressure limit. These previous observations are qualitatively consistent with our own measurements of $k_{1a}$ and $k_{1b}$. Since $k_{1b}$ is constant within the pressure range studied, it represents a constant offset to the amplitude of $k_1$. Two measurements of the rate constant $k_1$ for OD+CO in $N_2$ bath gas have been reported by Paraskevopoulos et al.[9] and Golden et al.[6] at room temperature. The comparisons are shown in Fig. 4. By calculating the quantity $k_1 = k_{1a}^{(N2)}[N_2]+k_{1b}$, the black solid line is obtained. Here, $k_{1a}^{(N2)}$ is the termolecular rate coefficient for *trans*-DOCO formation in $N_2$ bath gas measured by Bjork et al. The shaded teal region is the 1σ error for the calculated $k_1$. Because $k_{1a}$ was only measured to 75 torr, the vertical gray line is the demarcation of measured versus extrapolated regions of $k_1$. The good agreement with Paraskevopoulos et al. and Golden et al. provides quantitative validation for treating this OD+CO in the low-pressure limit as a simple superposition of the collision-induced combination reaction ($k_{1a}$) and chemical activated reaction ($k_{1b}$). As discussed by Fulle et al., one would expect that this treatment breaks down at much higher pressures upon the transition into the fall-off region, where more sophisticated modeling, e.g. Troe[32] corrections, RRKM[33] theory, etc., would be required. Finally, the *cis*-DOCO isomer may also have a non-negligible $k_{1a}$ contribution that has been unquantified to date.

Using the $k_{1a}$ reported by Bjork et al., we can also determine the branching yield for DOCO and D+$CO_2$ channels in $N_2$ and CO gas for the OD+CO reaction. Previous DOCO yield calculations[21] were made under the assumption that $k_{1b} = k_1$ at zero pressure by using an averaged $k_1$ value from Paraskevopoulos et al., Golden et al., and Westerberg et al.[34] We can now check the validity of this assumption by using the measured $k_{1b}$ in this work. The branching yield of DOCO is given by $k_{1a}/(k_{1a}+k_{1b})$. At the highest pressure of 75 torr, the DOCO yield in $N_2$ gas is 27±11%. and the corresponding $D+CO_2$ yield is 73±16%. This measured DOCO yield in $N_2$ is equivalent within 1σ to those calculated from Bjork *et al.*, and it is now supported with direct experimental validation. For CO bath gas, the same measurement procedure gives yields of 47±10% and 53±7% for DOCO and $D+CO_2$, respectively.

## 4. Conclusion

In this work, the OD+CO product branching to $D+CO_2$ has been quantified within the low-pressure limit. In conjunction with our previous $k_{1a}$ work, these results provide experimental evidence for the detailed balance of OD+CO product branching, whose sum yields the observed literature measurement of $k_1$. This work demonstrates another realization of the potential of optical frequency combs for studying complex chemistry problems: For systems like OD+CO that involve multiple intermediates and products, the inherent flexibility of time-resolved direct frequency comb spectroscopy allows for a comprehensive, quantitative, and deterministic exploration of detailed reaction mechanisms. The applications of frequency combs for studying many other classes of chemical reactions will only continue to grow with improving high finesse mirror technology and comb light sources extending beyond the mid-IR.[35, 36]

## 5. Acknowledgements

We acknowledge financial support from AFOSR, DARPA SCOUT, NIST, and the Physics Frontier Center at JILA (NSF). T. Q. Bui and B. Spaun are supported by the National Research Council Research Associate Fellowship, P. B. Changala is supported by the NSF GRFP, and O.H.

Heckl is partially supported through a Humboldt Fellowship. Finally, we thank Professor Mitchio Okumura at Caltech for helpful discussions and insights on the OD+CO work.

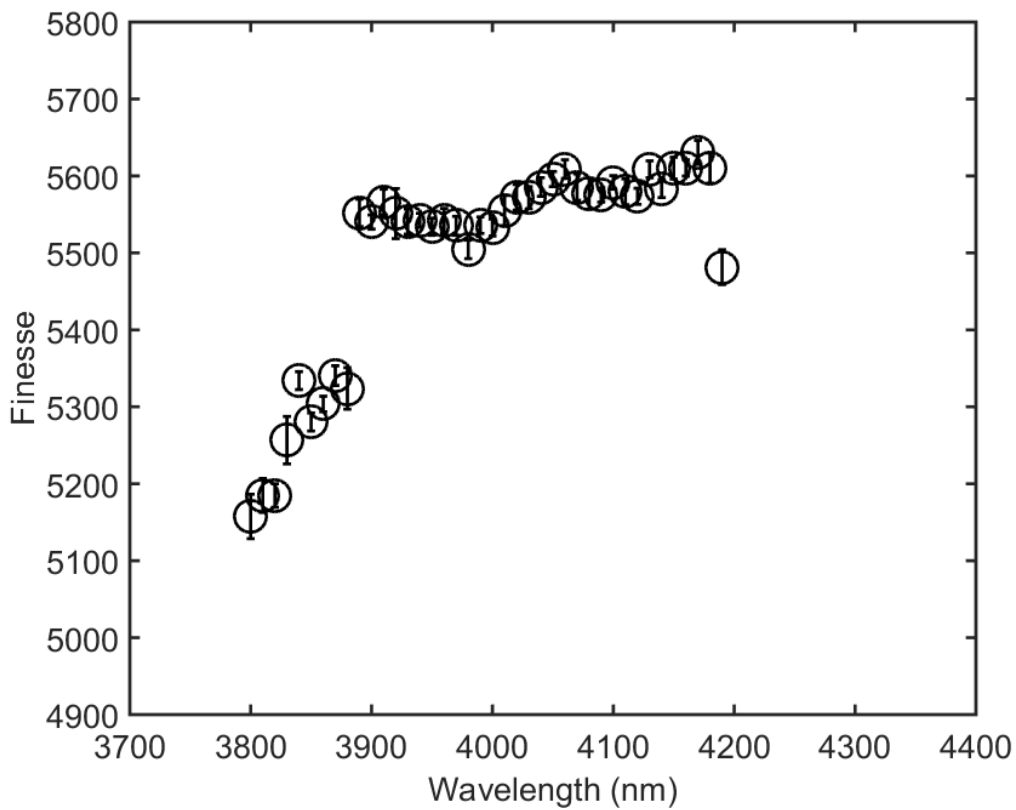

**Figure 1:** Measured finesse curve for high reflectivity mirrors centered at 3.92 μm. The large bandwidth (> 400 nm) provides access to most molecular species relevant to the OD+CO reaction, including *trans*-DOCO, $CO_2$, $DO_2$, $D_2O$, and OD(v=0-4).

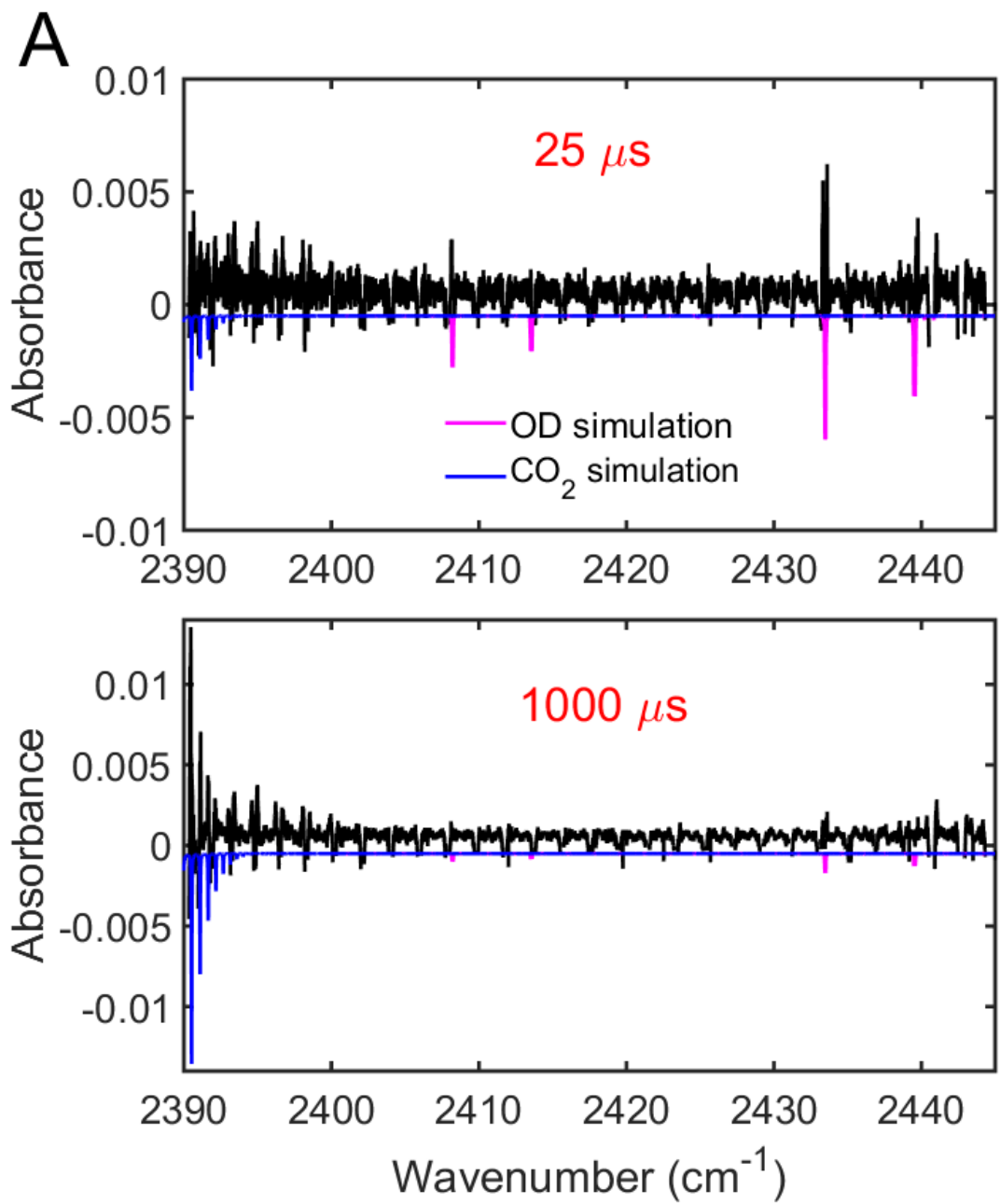

**Figure 2A.** Measured spectra (black) at 25 and 1000 μs delay from the photolysis pulse. These spectra are fitted to reference OD(v=0) (magenta) and $CO_2$ (blue) spectra to acquire the temporal profile. The decay of OD(v=0) and the rise of $CO_2$ are apparent between 25 and 1000 μs delay.

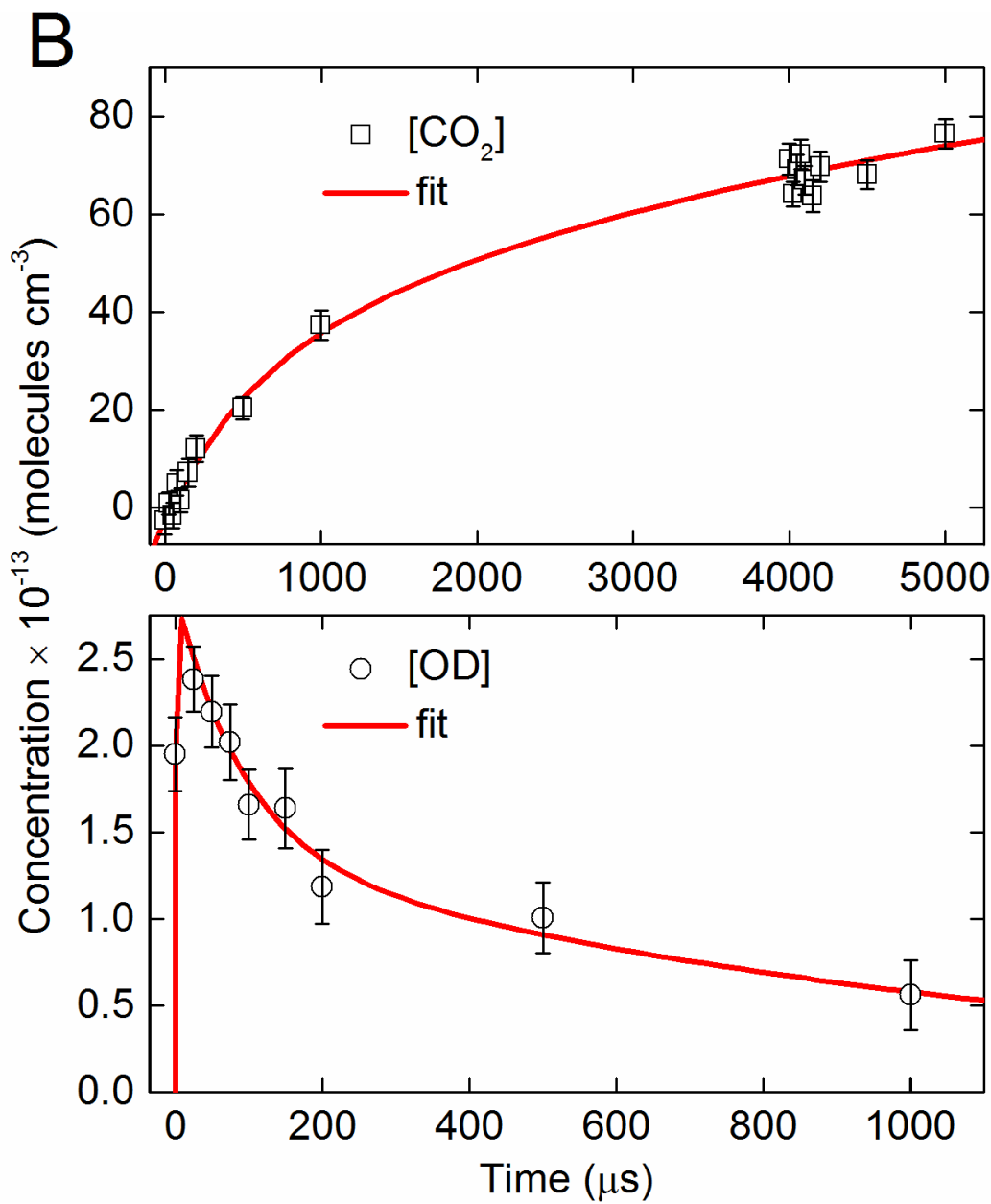

**Figure 2B.** (bottom) An analytical functional form for [OD]($t$) is obtained from fitting the data (black circles) to a sum of box-car averaged exponential rise and fall functions (red line). (top) The rise rate of CO is obtained from fitting the data (black squares) to eq. 6 (red line). The error bars correspond to 1$\sigma$ uncertainties, which include contributions from uncertainties from the spectral fits and concentration measurements. The camera integration time was fixed at 100 $\mu$s.

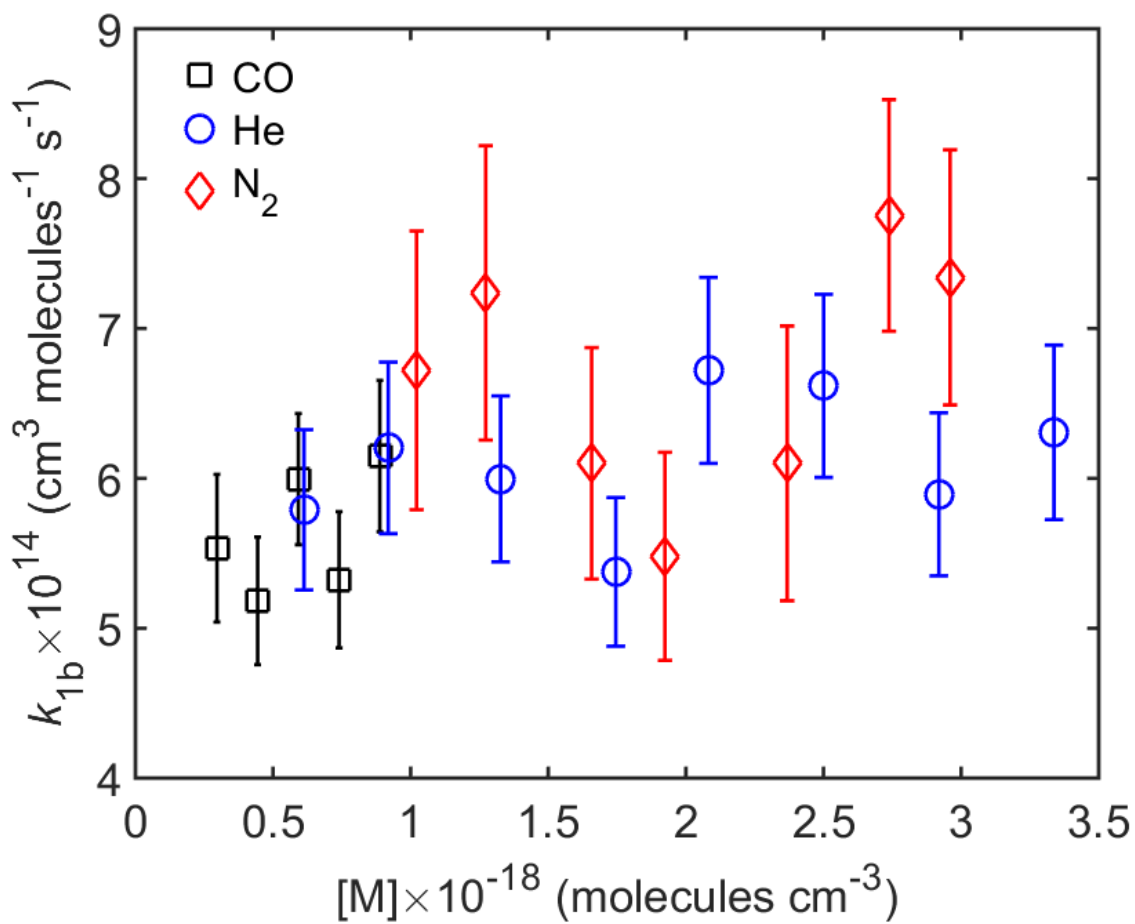

**Figure 3:** The formation rate ($k_{1b}$) of the activated products, D+$CO_2$, as a function of bath gases (M=$N_2$ (red diamond), CO (black squares), He (blue circles)) and pressure. $k_{1b}$ is calculated according to eq. 6. The error bars correspond to 1σ uncertainties, which include contributions from uncertainties from the fits to eq. 6 and concentration measurements.

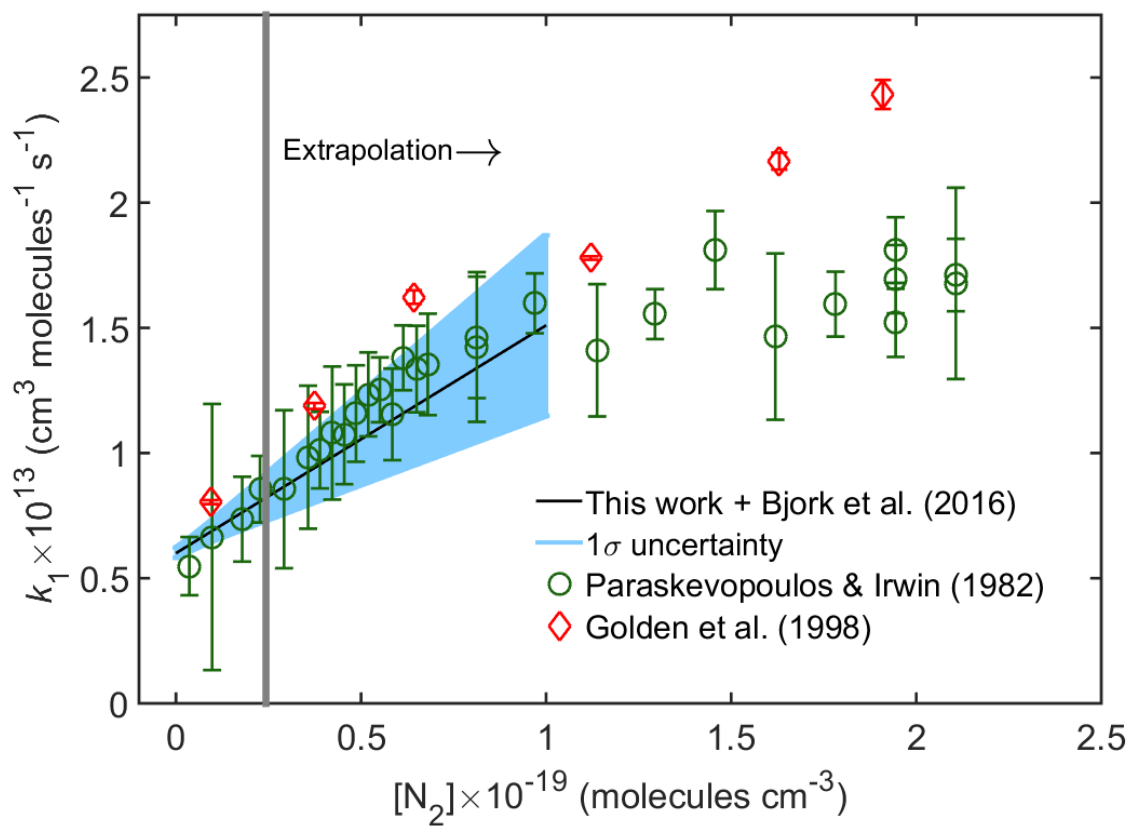

**Figure 4:** Comparison of the bimolecular rates, $k_1([M],T)=k_{1a}([M],T)+k_{1b}([M],T)$, for OD+CO in $N_2$ bath gas at room temperature. The solid black line is the calculated sum of $k_{1a}$ from Bjork et al. and $k_{1b}$ from this work with its 1σ uncertainty in shaded teal. $k_1$ for OD+CO in $N_2$ have been reported by Paraskevopoulos & Irvin (green circles) and Golden et al. (red diamonds). The gray vertical line marks the divide between the experimental (left) and extrapolated (right) pressure regions.